# SOLSAT: An FPGA Based satellite for effective Solar-activity monitoring

Dr Reena Monica P
SENSE(*School of Electronics*)
Vellore Institute of Technology.
Chennai Campus
Tamil Nadu, India
reenamonica@vit.ac.in

Ayushman Dash
SENSE(*School of Electronics*)
Vellore Institute of Technology.
Chennai Campus
Tamil Nadu, India
ayushman.dash2020@vitstudent.ac.in

*Abstract*— Kp-Index is a very important factor for determining how the Sun's magnetic field is affecting the Earth's magnetic field and help us prepare for the worst (like solar or geomagnetic storms). Currently there are around thirteen observation centers to determine the worldwide Kp index. Huge data is generated from all the observation centers which is then used to determine several factors, but the observatories are concentrated in places like North America and Europe only. To eliminate the data dependency from a very limited geographical periphery we can use a constellation of satellites in LEO which can log in data continuously and provide a real time and accurate Kp index which can help humanity for tackling with the impending issues from our nearest star in a better way. The constellation of the satellites will be deployed in the Low Earth Orbit(LEO) and continuously log the geomagnetic data and provide a constant stream of data, which will be more diverse given the area coverage made by one satellite. This is essentially a boon as more the data we can have better geomagnetic data and the Kp index will be all encompassing which can help in better mapping of any solar-catastrophe.

**Keywords—Kp-Index, Sun's magnetic field, Earth's magnetic field, geomagnetic storms, observation centers, global Kp-index, LEO satellites, real-time data, geomagnetic data, solar catastrophe**

## I. INTRODUCTION

The Kp-index plays a critical role in understanding how the Sun's magnetic field disrupts Earth's magnetic field, potentially leading to geomagnetic storms. These storms can have significant consequences for infrastructure and communication systems. Currently, the Kp-index is derived from data collected by a network of observatories concentrated in North America and Europe. This project proposes a novel approach to Kp-index determination: a constellation of satellites in Low Earth Orbit (LEO). First, it will eliminate the data dependency on a limited geographical region. Second, continuous data collection from multiple satellites will provide a more real-time and accurate picture of the Kp-index. Finally, the broader spatial coverage of the constellation will lead to a more comprehensive understanding of geomagnetic activity, enabling better preparation for solar events that could impact our planet. By providing a more robust and geographically diverse dataset, this LEO satellite constellation has the potential to significantly improve our ability to predict and mitigate the effects of solar storms on Earth's infrastructure and communication systems.

## II. LITERATURE SURVEY

The Kp index is a widely used measure of Earth's magnetic activity. This paper explains how the GFZ German Research Centre calculates and distributes this data.[1] The Kp index has been around for over 70 years and is crucial for space weather research and services. The authors describe both the historical and current methods for calculating the Kp index, including a recently improved technique for providing near real-time forecasts. They also acknowledge a slight bias in the data over time and estimate its impact. Overall, this paper offers valuable information for anyone using the Kp index.[8] Another research paper addresses the challenges posed by testing safety-critical autopilot systems for unmanned aerial vehicles (UAVs).[7] These systems require high reliability and safety standards, but extensive outdoor flight tests during development are costly and time-consuming. To tackle this, the paper introduces an indoor automatic test platform. Key aims include enhancing development efficiency and safety levels.[6] The platform comprises a unified modeling framework for various aerial vehicles, a real-time simulation platform (generated through automatic code and FPGA-based hardware-in-the-loop simulation), and an automatic test framework. The verification process validates the simulation platform's accuracy and credibility. Successful applications on multicopters demonstrate the practicality of this approach, emphasizing compatibility with future safety assessment and certification standards.[2] Yet another study explored calm periods in Earth's magnetic field (Quiet Periods) marked by low Kp-index values. They found a connection between the length of these periods and how often they occur. Interestingly, stricter limits for defining a Quiet Period led to a slower decline in their frequency.[3] However, a puzzling difference arose when analyzing daily sums versus individual Kp readings. The authors propose that averaging daily data might smooth out variations, leading to the discrepancy. While the cause of this difference needs further investigation, this research provides probability distributions for Quiet Period lengths, which could aid future predictions of both calm and active geomagnetic phases. The next step is to understand the physical reasons behind these patterns in the Earth's magnetic field.[10]The research paper titled "Flying Laptop: Operational Concept and On-Board Payload Data Processing" introduces the Flying Laptop, a small satellite developed by the Institute of Space Systems at the Universität Stuttgart. This satellite serves as a test bed for an on-board computer equipped with reconfigurable, redundant, and self-controlling capabilities based on field programmable gate arrays (FPGAs).[4]It's aims and goals include[6]: Efficiency Enhancement: The paper aims to significantly improve the development speed and safety level of unmanned vehicle autopilot systems.Unified Modeling Framework: It proposes a unified modeling framework for different types of aerial vehicles, facilitating shared modeling experience and failure mode analysis.Real-Time Simulation Platform: The development includes a real-time simulation platform generated through automatic code generation and FPGA-based hardware-in-the-loop simulation. This ensures simulation credibility on both software and hardware levels.Automatic Test Framework: The paper introduces an automatic test framework that traverses test cases during real-time flight simulation and assesses results. Some challenges however included[4, 3]: Cost and Time: Traditional outdoor flight tests are expensive and time-consuming.Safety: Ensuring safety during autopilot system development.The practical applications described here include[4, 7]:Successful applications on multicopters demonstrate the practicability of the proposed platform.Another relevant paper introduces a reconfigurable architecture for an on-board processor in space exploration systems.[5] This architecture leverages a dynamically reconfigurable multi-accelerator hardware design, ensuring transparent reconfiguration, scalability, dependability, and efficient power consumption at runtime. Integrated with the RTEMS operating system, it manages reconfiguration and fault mitigation while adhering to space requirements.[9] The proposed processor is applied to create a vision-based navigation system, enabling autonomous adaptation during different mission phases. Notably, reconfigurability allows practical use of Commercial Off-The-Shelf (COTS) Multiprocessor Systems-on-Chips (MPSoCs) in real-world scenarios. The architecture's features include in-flight adaptation, fault repair, and inherent fault tolerance, ensuring platform survival throughout the mission.[10, 7] Additionally, the RTEMS real-time operating system coordinates software tasks effectively.

Scientists analyzed current methods for predicting the Kp-index, which measures Earth's magnetic activity.[6] They tested how well these methods worked for short-term (half-day) and long-term (over 2 days) forecasts. The study found that short-term forecasts based on solar wind data were slightly improved by adding historical Kp values.[8] For long-term forecasts, however, historical patterns (recurrence) were the most accurate predictor, with solar wind data offering little benefit. The accuracy of predictions during high-activity periods was lower due to a lack of historical data for such events. Rebalancing the data improved forecasts for these periods. Overall, highly accurate predictions were only possible for short timeframes (6-20 hours). The authors suggest that entirely new approaches or additional data, like global solar observations, might be needed for truly long-term forecasts. They also propose using probability predictions instead of single values for longer horizons, which could help assess risks for various stakeholders.[11]

III. METHODOLOGY

## 3.1 FPGA

An FPGA (Field-Programmable Gate Array) board is a versatile electronic circuit board containing a reconfigurable logic core. This core, unlike a traditional processor, can be programmed to perform specific functions. The board typically includes memory for storing the FPGA configuration, input/output ports for connecting to sensors and actuators, and clocking circuitry to manage signal timing.

For a satellite application, the FPGA on the board can be programmed for real-time tasks like image compression, signal processing for scientific instruments, or even attitude control algorithms. This flexibility allows for in-flight updates or mission changes without requiring hardware modifications, making them a valuable tool for spaceborne applications. Keep your text and graphic files separate until after the text has been formatted and styled. Do not use hard tabs, and limit use of hard returns to only one return at the end of a paragraph. Do not add any kind of pagination anywhere in the paper. Do not number text heads-the template will do that for you.

### 3.1.1 FPGA Architecture

At its core, an FPGA board is a blank canvas for digital circuitry. It offers a reconfigurable logic architecture that can be programmed to perform specific functions. Here's a breakdown of its key components:

1. Configurable Logic Blocks (CLBs): These are the building blocks, containing elements like lookup tables (LUTs) and flip-flops. By configuring these elements, you can define the logic operations performed within the FPGA

2. Programmable Interconnect: This network acts like a highway system, allowing signals to travel between different CLBs. Switches within this network determine the specific routing paths based on the programmed configuration.

3. Input/Output (I/O) Blocks: These act as bridges, allowing the FPGA to communicate with external devices like sensors, actuators, and communication channels. They handle signal conditioning and buffering for proper data exchange.

4. Memory: This stores the configuration data (bitstream) that tells the FPGA how to connect its internal components (CLBs and interconnect). This essentially defines the functionality of the programmed circuit.

5. Clocking Circuitry: This generates and distributes clock signals that synchronize operations within the FPGA. Precise timing is crucial for reliable digital circuit operation.

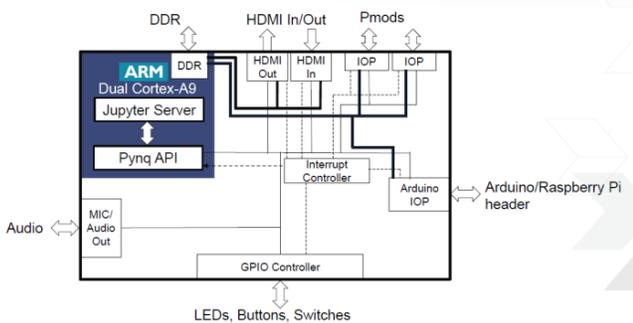

Fig. 1. The base architecture of an FPGA

### 3.1.2 FPGA in Space

The harsh environment of space demands adaptable and reliable electronics. This is where FPGA (Field-Programmable Gate Array) boards excel in satellite applications. Unlike traditional processors with fixed functions, FPGAs can be programmed to perform specific tasks needed on the mission. This reconfigurability allows for in-flight updates or mission changes without hardware swaps, a critical advantage for long-duration space missions

FPGAs also shine in real-time processing. Their ability to handle data streams efficiently is ideal for satellites that need to process sensor data from scientific instruments, control critical functions like attitude control, or manage high-speed communications in realtime. Furthermore, some FPGAs are designed to withstand radiation exposure, making them a reliable choice for the unforgiving space environment.

## 3.2 Process Work Flow

An HMC5883L magnetometer sends the local magnetic field then send the data to the FPGA (XILINX PYNQ-Z2) board which then processes the data into a Gauss unit, then it is converted into respective K-index which then is converted into Kp index per three hours in the board.

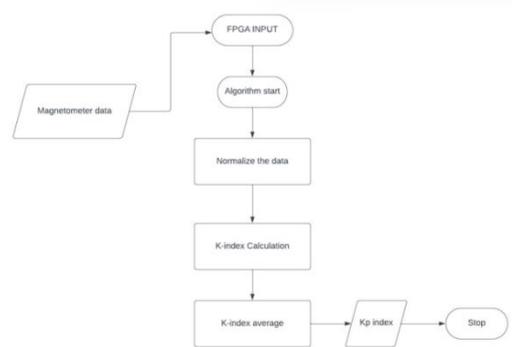

Fig. 2. A simple work-flow diagram

### 3.2.1 Xilinx Pynq-Z2 Architecture

The PYNQ-Z2 is an FPGA development board built around the Xilinx Zynq-7000 SoC (System on Chip). This SoC combines a dual-core Arm processor with programmable FPGA fabric. Users program the FPGA using Python through the PYNQ framework. The board also features Ethernet, HDMI, audio in/out, GPIO pins, and interfaces for Arduino and Raspberry Pi expansion modules. This allows for versatile embedded system development.

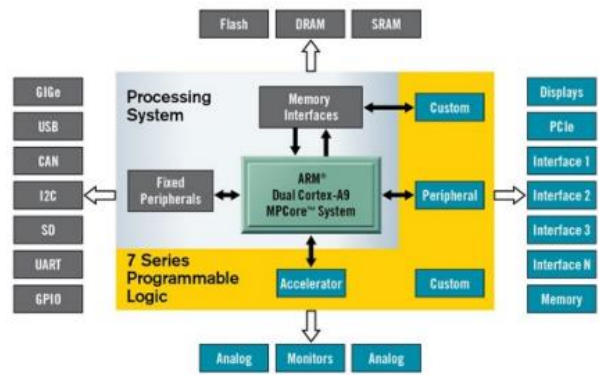

Fig. 3. PYNQ-Z2 Architecture

### 3.2.2 Proposed Working Procedure

The PYNQ-Z2's programmable FPGA fabric offers a powerful platform for interfacing with the HMC5883L magnetometer chip to measure the Earth's local magnetic field. By connecting the HMC5883L's I2C communication lines to the PYNQ-Z2's dedicated I2C pins, users can establish communication between the two devices. PYNQ's Python libraries can then be leveraged to write code that sends control commands to the HMC5883L, retrieves raw sensor data, and performs calibration to account for any sensor biases. With proper calibration, the magnetometer's data can be converted to meaningful magnetic field readings, providing valuable insights into the Earth's local magnetic field at the board's location.

### 3.2.3 HMC5883L chip

The HMC5883L is a tiny, low-power magnetometer chip designed to sense the Earth's magnetic field. This 3-axis magnetic sensor utilizes solid-state Anisotropic Magnetoresistive (AMR) technology for accurate and directional magnetic field measurements. It communicates digitally via I2C, making it easy to integrate with microcontrollers like the PYNQ-Z2. With a 1° to 2° compass heading accuracy, this chip is ideal for low-cost compassing and magnetometry applications.

*Fig. 4. Proposed Working Architecture Diagram*

## 3.3 Kp Index

The Kp-index is a global measure of geomagnetic activity, ranging from 0 (quiet) to 9 (extreme storms). It reflects disturbances in Earth's magnetic field caused by solar wind particles. Higher Kp values indicate stronger auroras, potential radio signal disruptions, and effects on power grids. Monitoring the Kp-index is crucial for understanding space weather and its impact on various technological systems.

### 3.3.1 Kp Index Concepts

The Kp-index is a vital tool for gauging geomagnetic activity, a measure of disturbances in Earth's magnetic field caused by charged particles from the sun's solar wind. Here's a breakdown of the Kp-index in detail:

1. Function: The Kp-index acts as a quantitative indicator, ranging from 0 (very quiet) to 9 (extreme storms), signifying the intensity of geomagnetic activity.

2. Data Acquisition: Ground-based magnetometers across the globe continuously measure fluctuations in the Earth's magnetic field. These measurements are then converted into local K-indices representing activity levels at each station over a 3-hour interval.

3. Global Kp-Index: The planetary Kp-index is calculated by combining the K-indices from a network of magnetometers, typically located at mid-latitudes. This provides a global picture of geomagnetic activity.

4. Impacts of Kp-Index: Higher Kp-values correlate with stronger auroras visible at higher latitudes. They can also disrupt radio communication by affecting the ionosphere, the layer of Earth's atmosphere that reflects radio waves. Additionally, extreme geomagnetic storms (Kp 5 and above) can induce electrical currents in power grids, potentially causing outages.

## 3.4 Kp Conversion

Magnetometers around the world capture Earth's magnetic field wiggles. These raw measurements are converted into local K-indices that reflect activity levels every 3 hours. Scientists then combine K-indices from a network of stations to create a single Kpindex, giving a global snapshot of geomagnetic storm intensity. This conversion process essentially boils down complex magnetic field data into a single, easy-to-understand number.

### 3.4.1 Location dependency

While the Kp-index is a global value, it relies on local data. Magnetometers like those in Ottawa, Canada or Canberra, Australia measure Earth's magnetic fidgets at their specific locations. These readings are then transformed into K-indices that reflect local activity. The final Kp-index combines K-indices from various observatories, offering a global view, but its foundation lies in these location-specific measurements.

### 3.4.2 Forecasting Table

There is a location based table that is divided into three groups depending on the magnetic activity of the Earth at that time of the year. The division is as follows:

1. JFND: January, February, November and December

2. MASO: March, April, September and October

3. MJJA: May, June, July and August

| | JFND | | | | | | | | MASO | | | | | | | | MJJA | | | | | | | |
|---|---|---|---|---|---|---|---|---|---|---|---|---|---|---|---|---|---|---|---|---|---|---|---|---|
| 0 | 0 | 0 | 0 | 0 | 0 | 0 | 0 | 0 | 0 | 0 | 0 | 0 | 0 | 0 | 0 | 0 | 0 | 0 | 0 | 0 | 0 | 0 | 0 | 0 |
| 1 | 0 | 4 | 4 | 3 | 3 | 3 | 3 | 3 | 4 | 4 | 3 | 2 | 3 | 3 | 3 | 3 | 3 | 3 | 3 | 2 | 2 | 2 | 3 | 3 |
| 2 | 7 | 8 | 7 | 7 | 6 | 6 | 5 | 5 | 6 | 7 | 6 | 6 | 6 | 5 | 5 | 5 | 6 | 5 | 4 | 4 | 5 | 6 | 6 | 6 |
| 3 | 9 | 11 | 11 | 11 | 10 | 8 | 7 | 8 | 9 | 11 | 11 | 10 | 9 | 7 | 7 | 8 | 9 | 10 | 8 | 7 | 7 | 9 | 11 | 11 |
| 4 | 12 | 14 | 13 | 12 | 11 | 10 | 10 | 11 | 11 | 13 | 14 | 12 | 11 | 11 | 11 | 12 | 12 | 14 | 13 | 10 | 11 | 13 | 15 | 16 |
| 5 | 15 | 17 | 19 | 17 | 16 | 15 | 13 | 14 | 16 | 17 | 19 | 18 | 16 | 14 | 13 | 14 | 16 | 16 | 17 | 15 | 15 | 15 | 15 | 16 |
| 6 | 18 | 20 | 22 | 21 | 20 | 18 | 17 | 16 | 19 | 21 | 23 | 21 | 20 | 20 | 20 | 20 | 18 | 20 | 21 | 21 | 17 | 18 | 18 | 20 |
| 7 | 21 | 23 | 25 | 24 | 22 | 21 | 20 | 19 | 23 | 25 | 25 | 24 | 22 | 22 | 22 | 22 | 21 | 22 | 23 | 22 | 21 | 21 | 21 | 22 |
| 8 | 24 | 25 | 26 | 26 | 24 | 22 | 22 | 23 | 23 | 25 | 25 | 24 | 23 | 22 | 22 | 23 | 23 | 23 | 23 | 23 | 23 | 23 | 23 | 23 |
| 9 | 27 | 27 | 27 | 27 | 27 | 27 | 27 | 27 | 27 | 27 | 27 | 27 | 27 | 27 | 27 | 27 | 27 | 27 | 27 | 27 | 27 | 27 | 27 | 27 |

### 3.5 Definition and Description of Technologies used

1. For programming the FPGA board we use Jupyter Notebook.

2. For conversion coding we use Colab notebook .

3. For connections we use the Raspberry-pi ports of the PYNQ-board, as the sensor uses the I2C protocol for communication.

NOTE: Not all of the technologies listed here were neither fully realized or implemented, due to limitations in available hands-on technology. We predominantly use PYNQ-Z2 board and Python for scripting and programming.

### 3.6 FPGA-board

*Definition*

An FPGA board is a programmable circuit board with building blocks you can arrange to do custom tasks.

#### 3.6.1 *Requirements*

PYNQ Image: The PYNQ-image is like a pre-configured operating system for your PYNQ-Z2 board. It's based on Linux and includes all the necessary tools and libraries to program the FPGA using Python. Think of it as a ready-made toolbox that lets you get started with your hardware experiments right away.

Micro-SD card: A micro-SD card with minimum of 8GB required to write and store the codes and most importantly host the PYNQ interface with PYNQ image.

UART: In the PYNQ-Z2, UART acts as a serial communication bridge. It allows your computer to talk to the board using a USB cable, similar to how you might connect to a printer. This is especially helpful for initial setup or accessing the terminal for troubleshooting.

#### 3.6.2 *Overview of PYNQ-image*

The PYNQ-Z2 relies on a special file called a PYNQ image. This image acts like a pre-programmed SD card that boots up the board. Built on a Linux foundation, the PYNQ image comes pre-loaded with all the necessary tools and libraries. These tools allow you to program the PYNQZ2's built-in FPGA (Field-Programmable Gate Array) using Python. Essentially, the PYNQ image provides a ready-to-use environment specifically designed to bridge the gap between Python and hardware development on the PYNQ-Z2 board.

Linux Kernel: This core provides a foundation for running other software components.

Python Environment: Python is the programming language you'll use to interact with the board's hardware. The PYNQ-image includes pre-installed libraries specifically designed for this purpose.

Device Drivers: These act as translators, allowing the Python environment to communicate effectively with the PYNQ-Z2's hardware components like the FPGA fabric.

Jupyter Notebook (Optional): This popular web-based interface is often included in the PYNQ-image. It provides a user-friendly platform to write, run, and visualize your Python code while interacting with the board.

#### 3.6.3 *Overview of HMC5883L*

The HMC5883L is a tiny but powerful 3-axis magnetometer, also known as a digital compass. It's like a miniature compass chip that senses the Earth's magnetic field in three directions. Here's what's inside:

Magneto-resistive Sensors: These act as the core, using special materials that change electrical resistance based on the magnetic field they experience.

Application Specific Circuit (ASIC): This chip amplifies the weak signals from the sensors, corrects for any biases, and converts them into a digital format for easy communication.

12-Bit Analog-to-Digital Converter (ADC): This component transforms the analog voltage signals from the ASIC into digital readings, allowing precise measurement of the magnetic field.

I2C Interface: This built-in communication protocol simplifies connecting the HMC5883L to other devices, like the PYNQ-Z2 board, for data exchange.

#### 3.6.4 *Sample code to observe the Serial Data Output*

This section contains the Python code to get the serial data from the Magnetometer sensor and display it continuously.

```python
def main():
    """
    Continuously reads magnetometer data, calculates heading, and prints results.
    """
    setup()

    while True:
        x = read_raw_data(X_MSB)
        y = read_raw_data(Y_MSB)
        z = read_raw_data(Z_MSB)

        heading = compute_heading(x, y)

        print(f"X: {x} uT, Y: {y} uT, Z: {z} uT, Heading: {heading:.2f}")
        time.sleep(0.5)

if __name__ == "__main__":
    main()
```

*Fig. 5. Python Code Snippet*

### 3.7 Sample Serial Output

The Python code that is mentioned above yields the following data in the output screen which can then be used to derive the local magnetic field

*Fig. 6. Sample Output*

### 3.8 Conversion Code

This code sample displays the conversion of raw data from the HMC5883L magnetometer via using the conversion and forecasting table. This code takes in the continuous data and measures the Kp index out of it.

*Fig. 7. Conversion Sample*

## IV. RESULTS AND DISCUSSIONS

### A. *Objective*

Here we are simply trying to measure the local magnetic field of the Earth via the HMC5883L sensor and log the output with the help of the FPGA Xilinx PYNQ-Z2 board this will then allow us to process the data combined with some additional factors like the geo location, and the date and time, which will allow us to calculate and forecast the K index accurately which can then be copared with the ESA database. The results shown here are the K index of any given month with the hourly magnetic data of any given day of the month. The conversion method used here is strictly adhered by the Upsala observatory official conversion table.

### *4.1 Results and Comparison*

Below shown results are for all the three distinct groups of months for which Kp index is calculated

- JFND (January, February, November, December)
- MASO (March, April, September, October)
- MJJA (May, June, July, August)

In the following chart all the data is represented in the above order for the same magnetic dataset.

#### *4.1.1 K-index for JFND*

The bar chart shown below shows the local K-index for the months January, February, November or December.

*Fig. 8. January*

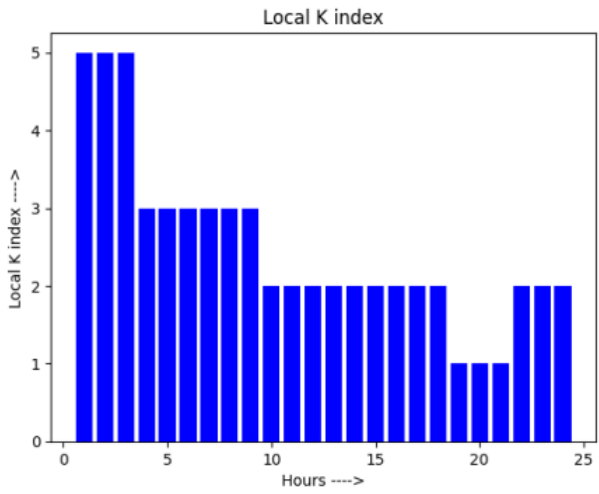

Fig. 9. February

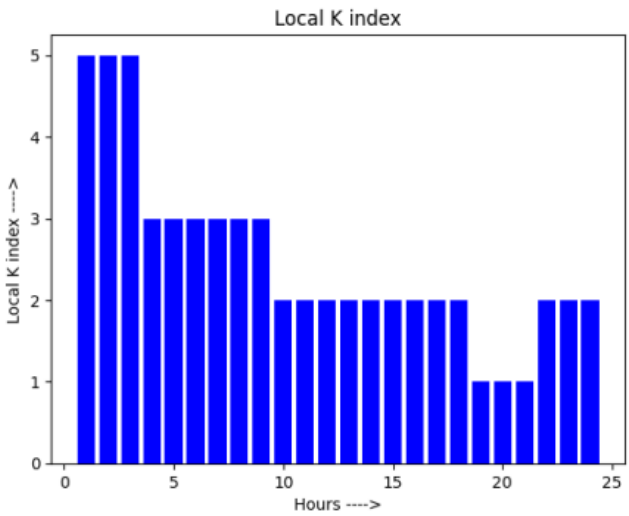

Fig.10. November

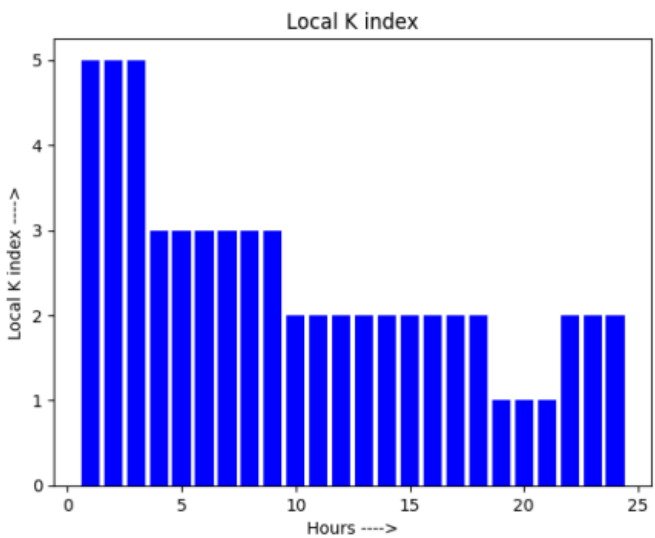

Fig.11. December

The bar chart shown below shows the local K-index for the months March, April, September and October.

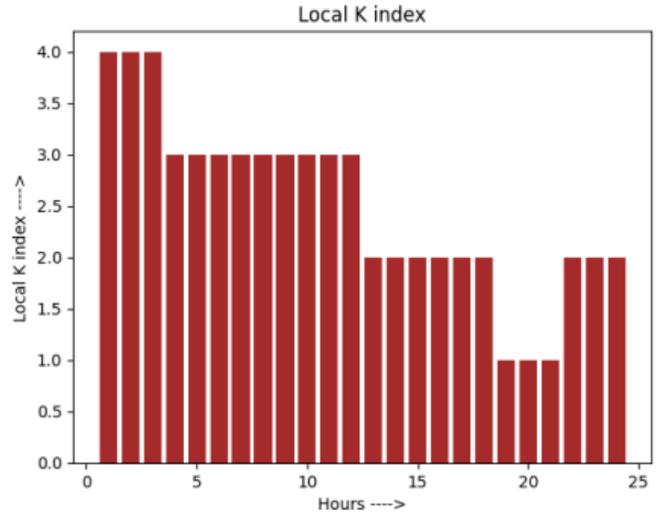

Fig.12. March

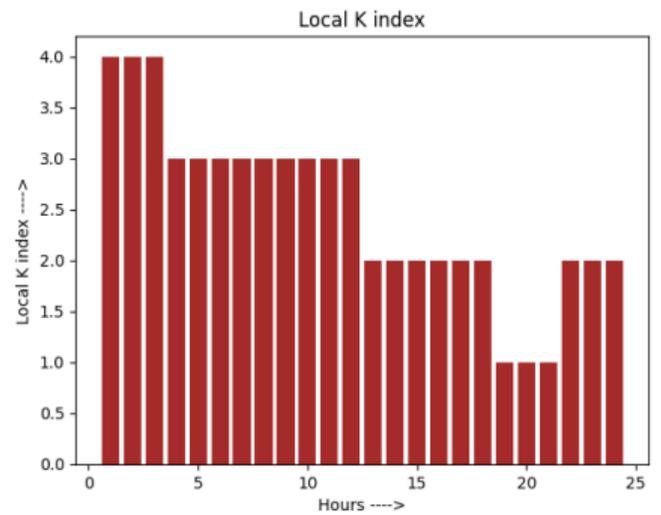

Fig.13. April

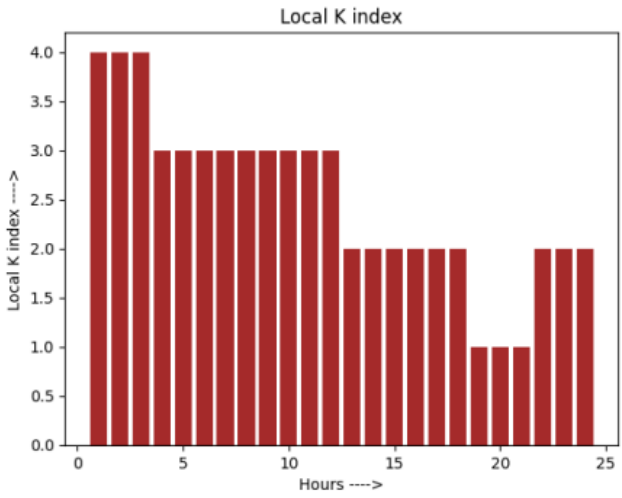

*Fig.14. September*

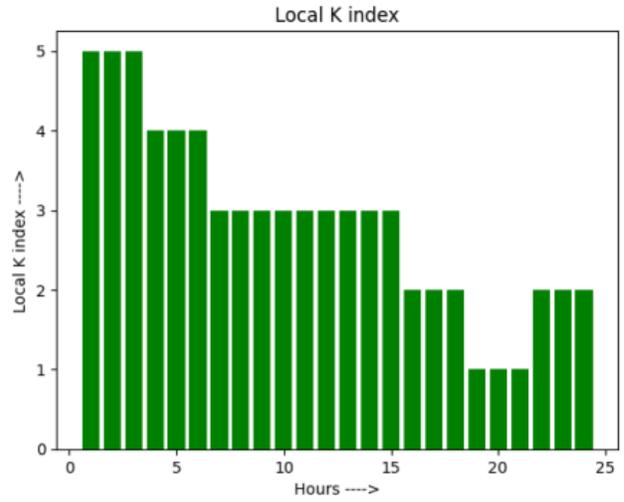

*Fig.16. May*

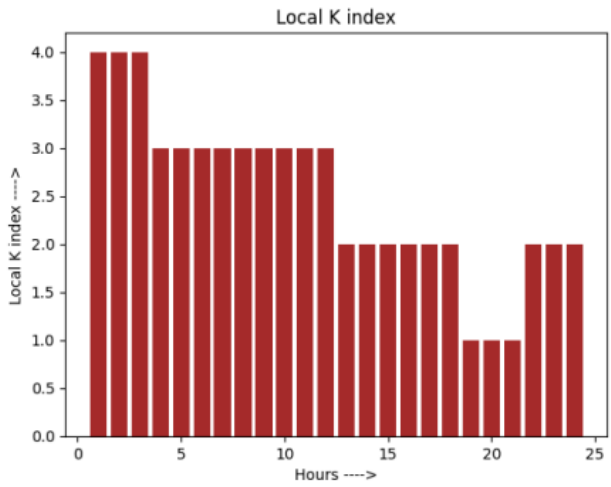

*Fig.15. October*

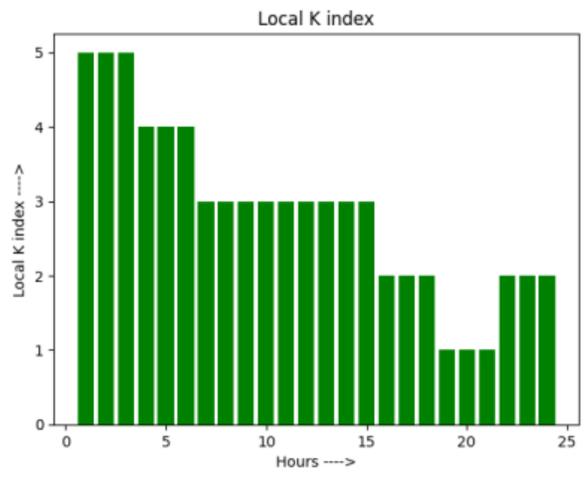

*Fig.17. June*

### 4.1.3    K-index of MJJA

The bar chart shown below shows the local K-index for the months May, June, July and August.

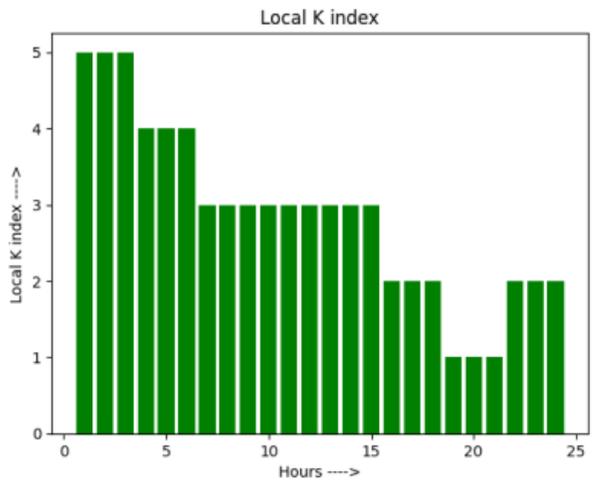

*Fig.18. July*

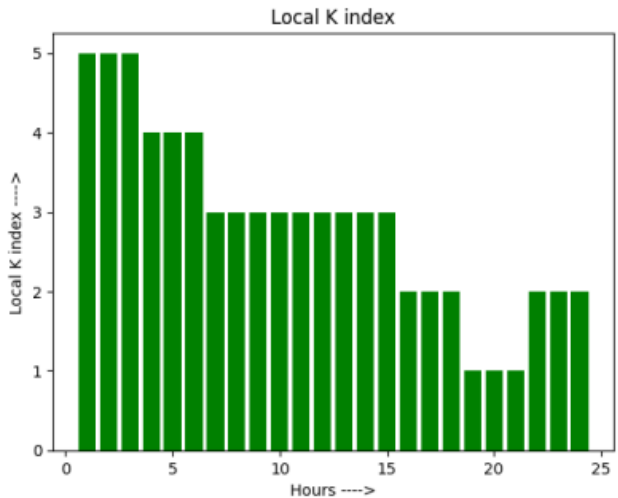

*Fig.19. August*

## V. CONCLUSION AND FUTURE SCOPE

A system similar to this had been planned by the ESA (Eurpean Space Agency). The mission name was called SWARM that was launched back in 22nd November 2013. It is a constellation of 3 EO (Earth Observatory) Satellites namely Alpha, Bravo and Charlie. It features a plethora of advances magnetometer and laser equipment for better observation.

A network of Earth observation satellites could revolutionize our understanding of the Earth's magnetic field. By taking measurements from multiple locations simultaneously, these satellites would create a much more detailed and comprehensive picture of the field's behavior. This enhanced data would improve our ability to track and predict fluctuations in the field, particularly those caused by incoming solar storms. With more accurate forecasts, we could take preventative measures to safeguard power grids, communication systems, and satellites from potentially damaging geomagnetic disturbances. This proactive approach could significantly reduce the risk of widespread blackouts, communication outages, and disruptions to critical infrastructure, ultimately mitigating the cascading effects of solar storms.

While the current Kp-index system relies on a network of roughly thirteen ground-based observatories, a constellation of Earth observation satellites promises a game-changer. These satellites would provide near-global, real-time magnetic field data, surpassing the limitations of geographically restricted observatories. This shift from a sparse network to a comprehensive satellite constellation holds immense potential for more accurate and spatially detailed measurements of geomagnetic activity

## ACKNOWLEDGMENT

I am grateful to Dr. Reena Monica P of School of Electronics Engineering at VIT Chennai, for her invaluable guidance, encouragement, and patience throughout this research. Her mentorship extended beyond academics, allowing me to learn from an expert in VLSI and Hardware Design.

## REFERENCES

[1] Xunhua Dai, Chenxu Ke, Quan Quan and Kai-Yuan Cai' RFlySim: Automatic test platform for UAV autopilot systems with FPGA-based hardware-in-the-loop simulations' School of Computer Science and Engineering, Central South University, Changsha 410000, China and School of Automation Science and Electrical Engineering, Beihang University, Beijing, 100191, China.

[2] Toshinori Kuwahara , Felix B¨ohringer , Albert Falke , Jens Eickhoff , Felix Huber, Hans-Peter R¨oser a 'Integration of Big Data Using Semantic Web Technologies.' Institute of Space Systems, Universit¨at Stuttgart, Pfaffenwaldring 31, 70569 Stuttgart, GermanybEADS Astrium GmbH, 88039 Friedrichshafen, GermanycSteinbeis Transferzentrum Raumfahrt, R¨otestr. 15, 71126 G¨aufelden, Germany

[3] ARTURO PEREZ 1 , ALFONSO RODRIGUEZ 1 , (Student Member, IEEE), ANDRES OTERO1 , (Member, IEEE), DAVID GONZ ´ALEZ ARJONA2 , ´ALVARO JIM ´ENEZ-PERALO2 , MIGUEL ´ANGEL VERDUGO2 , AND ED- ´ UARDO DE LA TORRE a 'Run-Time Reconfigurable MPSoC-Based On-Board Processor for Vision-Based Space Navigation' Centro de Electr´onica Industrial, Universidad Polit´ecnica de Madrid, 28006 Madrid, Spain 2GMV Aerospace and Defence SAU, 28760 Madrid, Spain

[4] Collette Walbeck. 'The Kp Index and Behavior of Quiet Periods.' PHYS 4900 Undergrad Research Department of Physic, Utah State University Dr. Jan Sojka April 21, 2023

[5] J. Matzka1, C. Stolle1,2, Y. Yamazaki1, O. Bronkalla1, and A. Morschhauser. 'The Geomagnetic Kp Index and Derived Indices of Geomagnetic Activity.' FGFZ German Research Centre for Geosciences, Potsdam, Germany, 2Faculty of Science, University of Potsdam, Potsdam, Germany

[6] S. Wing,1J. R. Johnson,2J. Jen,1C.-I. Meng,1D. G. Sibeck,3K. Bechtold,1J. Freeman,5K. Costello,6M. Balikhin,4and K. Takahashi 'Kp forecast models' Citation:Wing, S., J. R. Johnson, J. Jen, C.-I. Meng, D. G. Sibeck, K. Bechtold, J. 37 Bibliography. 38 Freeman, K. Costello, M. Balikhin, andK. Takahashi (2005), Kp forecast models,J. Geophys. Res.,110, A04203, doi:10.1029/2004JA010500.

[7] Yuri Y. Shprits1,2,3, Ruggero Vasile1, and Irina S. Zhelavskaya. 'Nowcasting and Predicting theKpIndex Using HistoricalValues and Real-Time Observations' 1GFZ German Research Centre for Geosciences, Potsdam, Germany,2Institute of Physics and Astronomy, University ofPotsdam, Potsdam, Germany,3Department of Earth, Planetary, and Space Sciences, University of California, LosAngeles, CA, USA.

[8] Heather A. Elliott,1J¨org-Micha Jahn,1and David J. McComas1'TheKpindex and solar wind speed relationship:Insights for improving space weather forecasts.' SPACE WEATHER, VOL. 11, 339–349, doi:10.1002/swe.20053, 2013

[9] Jingjing Wang1,2*, Bingxian Luo1,2,3, Siqing Liu1,2,3 and Liqin Shi 'A machine learning-based model for the next 3-day geomagnetic index (Kp) forecast' State Key Laboratory of Space Weather, National Space Science Center, Chinese Academy of Sciences, Beijing, China, Key Laboratory of Science and Technology on Environmental Space Situation Awareness, Chinese Academy of Sciences, Beijing, China, University of Chinese Academy of Sciences, Beijing, China

[10] GABRIEL MAIOLINI CAPEZ , SANTIAGO HENN CONICET -JUAN A. FRAIRE , Senior Member, IEEE CONICET Univ ROBERTO GARELLO 'Sparse Satellite Constellation Design for Global and Regional Direct-to-Satellite IoT Services.' Graduate Student Member,

IEEE Politecnico di Torino, Turin, Italy, Universidad Nacional de Córdoba, Córdoba, Argentina, Lyon, Inria, INSA Lyon, CITI, Villeurbanne, France Saarland University, Saarbrücken, Germany, Politecnico di Torino, Turin, Italyo.

[11] Erkan İnceöz, Raşit Tutgun, Ayşe Melda Yüksel Turgut. 'FPGA Based Transmitter Design Using Adaptive Coding and Modulation Schemes for Low Earth Orbit Satellite Communications.' Electrical and Electronics Engineering, TOBB University of Economics and Technology, Ankara, Turkey, Space Technologies Research Institute, The Scientific and Technological Research Council of Turkey, Ankara, Turkey.